\pdfoutput=1

\documentclass[11pt]{article}

\usepackage[preprint]{acl}

\usepackage{times}
\usepackage{latexsym}

\usepackage[T1]{fontenc}

\usepackage[utf8]{inputenc}

\usepackage{microtype}

\usepackage{inconsolata}

\usepackage{graphicx}
\usepackage{multirow}
\usepackage{graphics}
\usepackage{arydshln}

\usepackage{colortbl}
\usepackage[table]{xcolor}
\usepackage{diagbox}
\usepackage{tabularx}              
\usepackage{booktabs}              
\usepackage{array}                 
\usepackage{kotex}[cjk]
\usepackage{amsmath}
\usepackage{tcolorbox}
\usepackage{subcaption}
\usepackage{geometry}
\usepackage{amssymb}

\newcommand{\header}[1]{\vspace{2mm}\noindent\textbf{#1}}

\title{SemBridge: Language Transfer in Sparse Encoders via Multilingual Semantic Bridges}

\author{
 \textbf{Seongtae Hong\textsuperscript{1}},
 \textbf{Youngjoon Jang\textsuperscript{1}},
 \textbf{Jia-Huei Ju\textsuperscript{2}},
 \textbf{Hyeonseok Moon\textsuperscript{1}$^{*}$},
 \textbf{Heuiseok Lim\textsuperscript{1}\thanks{Corresponding author}}\\
 \textsuperscript{1}Department of Computer Science and Engineering, Korea University\\
 \textsuperscript{2}University of Amsterdam\\
 \texttt{\{ghdchlwls123,dew1701,glee889,limhseok\}@korea.ac.kr}\quad 
 \texttt{j.ju@uva.nl}
}

\begin{document}
\maketitle
\begin{abstract}
Sparse encoders offer high-precision retrieval by representing term importance within a vocabulary space, yet their English-centric structures pose a critical impediment to language transfer for non-English languages. To overcome this structural limitation, we propose SemBridge, a novel embedding initialization method designed for cross-lingual adaptation in sparse encoders by leveraging multilingual bridge models. SemBridge establishes semantic alignments between source and target vocabularies using multilingual dense embeddings as a bridge. Rather than directly relying on all source tokens, SemBridge selects a small set of semantically related source-language tokens and uses them to initialize each target-language token, effectively filtering out semantic noise and reconstructing target tokens as precise linear combinations of core synonyms. This accelerates convergence during fine-tuning and improves training efficiency. Extensive experiments across five languages and four sparse architectures demonstrate that SemBridge achieves superior zero-shot retrieval performance and consistently improves retrieval performance after fine-tuning compared to existing baselines. These results validate SemBridge as a practical solution for deploying high-performance sparse retrieval systems in diverse linguistic environments.
\end{abstract}

\section{Introduction}

Information Retrieval has evolved towards deep learning-based dense retrieval to address the lexical mismatch problem~\cite{mismatch, mismatch2, ance, cho, dpr, coCondenser}. To overcome dense retrieval's low interpretability and lack of explicit term-matching capabilities~\cite{inter}, sparse encoder models have emerged as an alternative~\cite{deepct, sparta, splade}. By representing the contextual importance of terms as sparse vectors within the vocabulary space, they achieve both semantic understanding and keyword precision. Furthermore, their direct compatibility with existing Inverted Index infrastructure significantly enhances the efficiency of large-scale systems~\cite{sparterm, inverted, efficiency, efficiency2}. These vocabulary-level representations also provide human-readable term weights, offering interpretable evidence for why a document is retrieved. As the demand for globalized information access grows, efforts to build language-specific retrieval models have gained increasing attention. While most of these efforts have centered on dense retrieval, extending sparse encoders to new linguistic environments is not straightforward. Unlike dense retrievers, whose representations are formed in continuous latent spaces, sparse encoders rely on their vocabulary space as the explicit output space for retrieval. Constrained by this inherent structure, merely fine-tuning an existing English-centric sparse encoder for a target language does not easily yield performance gains.

\begin{figure}
  \includegraphics[width=\columnwidth]{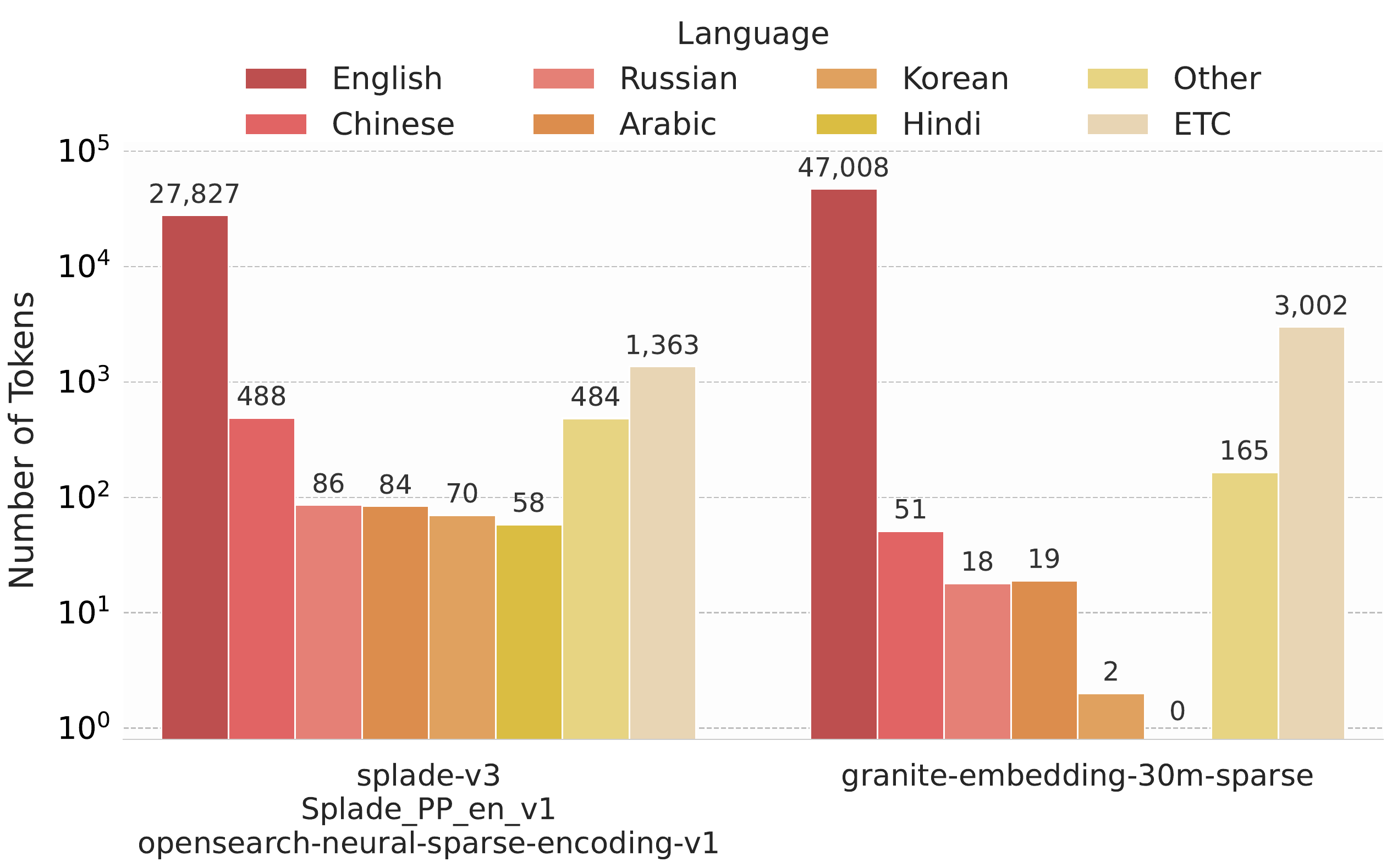}
  \caption{Token distribution across sparse encoders. \textit{Other} and \textit{ETC} represent additional language and non-linguistic tokens, respectively.}
  \label{fig:teaser}
\end{figure}

The underlying cause is evident from Figure~\ref{fig:teaser}, which provides an empirical analysis of vocabulary distributions across sparse encoder models. Our analysis demonstrates that for most models, the proportion of non-English tokens is negligible. Notably, the granite-30m-sparse contains only two Korean tokens, and splade-v3 similarly exhibits a significant bias toward English. Given that the vocabulary in a sparse encoder serves as the explicit output space for representing semantics, the absence of target-language tokens creates a structural lack of dimensions through which the model can assign importance to target-language terms. 
Consequently, it is intrinsically difficult to capture the nuanced semantics of non-English languages within such an English-centric structure, posing a critical bottleneck in reproducing the source model's retrieval capabilities in a target language. Therefore, even with target-language fine-tuning, the scarcity of target-language tokens remains a key limiting factor in achieving optimal performance.

In this paper, to overcome these structural limitations and effectively deploy sparse encoders in target language environments, we propose SemBridge, a novel embedding initialization method that preserves the existing capabilities of a source sparse encoder while transferring its source-language knowledge to a target language. We leverage multilingual dense embeddings as a bridge to perform token-level semantic alignment between the source and target language vocabularies. By reconstructing sophisticated semantic correspondences between tokens with differing surface forms within the parameter space, our method initializes target token embeddings that serve as an optimal starting point. Rather than assigning target tokens randomly or relying only on surface overlap, SemBridge initializes each target token by selecting semantically related source-language tokens and transferring their embedding information through sparse semantic weighting. This ensures that the source model's inherent retrieval capabilities are fully preserved and remain immediately effective in the target language environment.

To demonstrate the generalizability and utility of the proposed method, we conduct extensive experiments across four sparse models and five languages: Arabic, Chinese, Hindi, Korean, and Russian. Experimental results confirm that SemBridge effectively transfers the source model's retrieval capabilities in a zero-shot setting; furthermore, it achieves superior performance and faster convergence compared to baselines through fine-tuning. Through qualitative analysis, we further reveal that our method precisely aligns target language tokens with core synonyms in the source vocabulary while effectively filtering out unnecessary semantic noise. These results substantiate that SemBridge transcends lexical barriers to fully transplant the source model's semantic discernment into target language environments. Ultimately, SemBridge serves as a practical and efficient solution for adapting and building high-performance sparse retrieval models in target-language environments, even in non-English settings facing data scarcity.

\section{Related Work}
\subsection{Sparse Encoder}

Sparse encoders are first-stage retrieval models that represent text as high-dimensional sparse vectors by predicting token importance within the vocabulary space. Various approaches have been proposed to advance this paradigm, including learning the semantic importance distribution for all terms~\cite{sparterm}, re-estimating the weights of existing terms~\cite{deepct}, expanding indices by predicting latent terms~\cite{doc2query}, and maximizing token-level interactions~\cite{coil,sparta}. These approaches have garnered significant attention due to their practicality and interpretability. Because the encoded output aligns with the vocabulary, it can directly utilize existing Inverted Index infrastructure, enabling efficient retrieval without high-cost Approximate Nearest Neighbor (ANN) search indices~\cite{unicoil,sparseembed}. Furthermore, they explicitly reveal the tokens contributing to the retrieval score~\cite{splade} and allow flexible control over the balance between memory usage and performance~\cite{efficient,splade++}. However, many recent sparse encoders are predominantly trained on English~\cite{granite,spladepp}. In sparse encoders, where the vocabulary space itself serves as the representation space, a small proportion of target language tokens leads to a structural lack of ``dimensions'' to represent that language, making simple fine-tuning ineffective. While sparse encoders trained specifically for certain languages exist~\cite{spladefr,spladeko}, training such models requires a strong MLM model trained from scratch on the corresponding language, large-scale retrieval training data, and significant computational resources.

\subsection{Language Transfer}
Language transfer primarily refers to an approach that adapts models pre-trained in resource-rich languages, such as English, to a target language environment to efficiently achieve performance even with limited data and computational resources. Generally, transfer is attempted through continued pretraining and finetuning~\cite{lapt3, lapt1, lapt2}. Another line of work performs vocabulary expansion, which introduces new target language tokens into the existing vocabulary and initializes embeddings only for those added tokens~\cite{expansion1, expansion2}. Beyond vocabulary expansion, tokenizer replacement directly addresses vocabulary mismatch by replacing the source tokenizer with one constructed for the target language. In this setting, the central challenge is how to initialize the target tokenizer embeddings while preserving the representation space of the source model. Basic strategies, such as random or source-statistics-based initialization, preserve only generic or distributional properties and fail to align source and target token semantics~\cite{random,mean}. Prior work on semantic embedding initialization addresses this issue using bilingual lexical resources, auxiliary embedding spaces, or matrix factorization~\cite{wechsel,focus,ofa,trans-token}. However, these methods often rely on language-pair-specific lexicons, lexical overlap, or low-rank approximations, which can restrict the scope or fidelity of token level semantic transfer.

\section{SemBridge}
\begin{figure}
  \center
  \includegraphics[width=0.9\columnwidth]{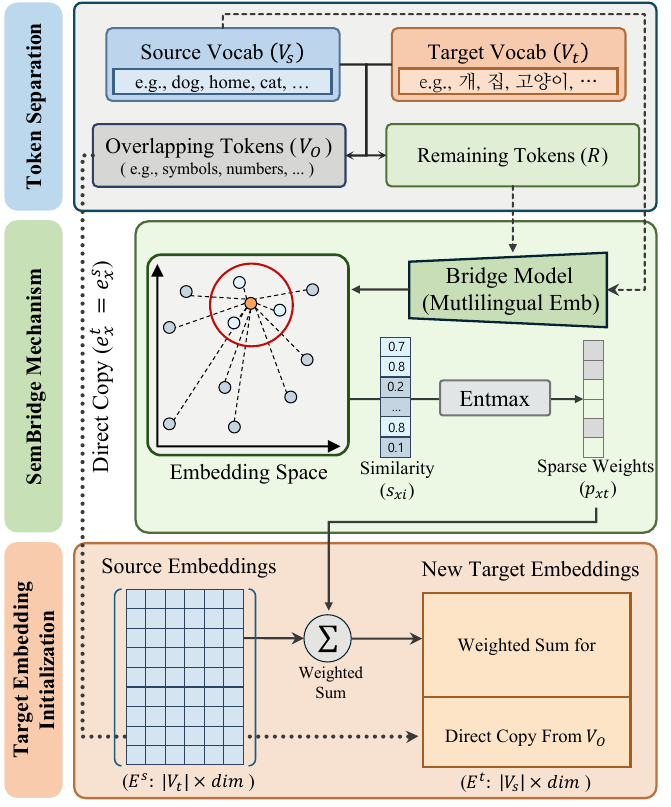}
  \caption{Overview of the SemBridge embedding initialization process.}
  \label{fig:methods}
\end{figure}

In this section, we introduce SemBridge, which leverages a source sparse encoder model $M$ to initialize embeddings tailored for a target language, as illustrated in Figure~\ref{fig:methods}. Let $T_{s}$ and $\mathcal{V}_{s}$ be the source tokenizer and vocabulary of $M$, and $T_{t}$ and $\mathcal{V}_{t}$ be the target tokenizer and vocabulary, respectively. We denote the embedding vector of a source token $x_s \in \mathcal{V}_{s}$ as $\mathbf{e}^{s}_{x_s} \in \mathbb{R}^d$, and the embedding vector to be initialized for a target token $x_t \in \mathcal{V}_{t}$ as $\mathbf{e}^{t}_{x_t}$.

\subsection{Overlapping Token Embedding Transfer}
\label{sec:overlap}
Although trained on different languages, source and target tokenizers often share language-agnostic tokens, such as numbers, symbols, or proper nouns. To leverage these shared tokens, we identify the overlapping token set $\mathcal{V}_{o} = \mathcal{V}_{t} \cap \mathcal{V}_{s}$. This set includes not only exact string matches but also tokens deemed identical after pre-processing normalization (e.g., ignoring case or whitespace). For any target token $x \in \mathcal{V}_{o}$, we initialize its embedding by directly copying the source token's embedding:
\begin{equation}
\mathbf{e}^{t}_{x} = \mathbf{e}^{s}_{x}, \quad \forall x \in \mathcal{V}_{o}
\end{equation}
This approach transfers the universal semantic information learned by the source model to the target model, and in particular, enhances the initial stability of the target model by preserving the representational expressiveness of the source model.

\subsection{Cross-lingual Semantic Bridge}
\label{sec:bridge}
The majority of tokens in the target vocabulary possess surface forms distinct from those of the source tokens, yet they remain semantically closely linked. To effectively transfer semantics across these fundamental lexical mismatches, they necessitate precise mapping within a semantic representation space. Accordingly, we employ a multilingual dense embedding model $\mathcal{B}$\footnote{Using bge-m3~\cite{bge} as the bridge model $\mathcal{B}$.} as a semantic bridge to project both source and target tokens into a shared vector space for semantic-based alignment.

Specifically, we define the set of remaining tokens to be newly initialized as $\mathcal{R} = \mathcal{V}_{t} \setminus \mathcal{V}_{o}$. Each source token $x_s \in \mathcal{V}_{s}$ and each target remaining token $x_t \in \mathcal{R}$ are fed into model $\mathcal{B}$ to obtain their corresponding dense representations, defined as:
\begin{equation}
\begin{aligned}
\mathbf{h}_{x_s} &= \mathcal{B}(x_s), \quad \forall x_s \in \mathcal{V}_{s} \\
\mathbf{h}_{x_t} &= \mathcal{B}(x_t), \quad \forall x_t \in \mathcal{R}
\end{aligned}
\end{equation}

Next, for each remaining token $x_t \in \mathcal{R}$, we calculate its similarity with all tokens in the source vocabulary $\mathcal{V}_{s}$. Let $N = |\mathcal{V}_{s}|$, and let the source tokens be denoted by $\{x_{s_1}, x_{s_2}, \dots, x_{s_N}\}$. The similarity vector $\mathbf{s}_{x_t} \in \mathbb{R}^{N}$ for an uninitialized target token $x_t$ is constructed as follows:

\begin{equation}
\mathbf{s}_{x_t} = \left[ \cos(\mathbf{h}_{x_t}, \mathbf{h}_{x_{s_1}}),  \cdots,  \cos(\mathbf{h}_{x_t}, \mathbf{h}_{x_{s_N}}) \right]
\end{equation}

By calculating these similarity vectors independently for all $M = |\mathcal{R}|$ remaining tokens, we obtain the final similarity matrix $\mathbf{S} \in \mathbb{R}^{M \times N}$:
\begin{equation}
\mathbf{S} =
\begin{bmatrix}
\mathbf{s}_{x_{t_1}} \\
\vdots \\
\mathbf{s}_{x_{t_M}}
\end{bmatrix},
\quad i = 1, \dots, M.
\end{equation}
Consequently, the matrix $\mathbf{S}$ quantifies the semantic relevance between the uninitialized target tokens and the entire source vocabulary. This plays a crucial role in deriving the weight vector for target token embedding initialization.

\subsection{Similarity-Based Sparse Weighting for Target Token Embedding Initialization}
\label{sec:sparse-weight-init}
To initialize the embedding for each remaining target token $x_t$, we transform its computed similarity vector $\mathbf{s}_{x_t}$ into a wight vector, which is then used to compute the weighted average of the source token embeddings. We calculate this vector by individually apply the Entmax~\cite{entmax} transformation to the similarity vector $\mathbf{s}_{x_t}$. The primary motivation for utilizing this specific transformation is the active removal of semantically irrelevant tokens that can cause unnecessary interference within the source vocabulary space. Entmax dynamically selects only a few highly relevant tokens by inherently blocking such noise through truncating the tail of the probability distribution to exact zeros. Specifically, the sparse weight vector $\mathbf{p}_{x_t}$ corresponding to each $x_t \in \mathcal{R}$ is calculated as follows:
\begin{equation}
\label{eq:sparse_weighting}
\mathbf{p}_{x_t} = \mathrm{Entmax}_{\alpha}(\mathbf{s}_{x_t}),
\; \text{s.t.} \; \sum\limits_{i=1}^{N} p_{x_{ti}} = 1.
\end{equation}
The hyperparameter $\alpha$ governs the degree of sparsity in the resulting weight vector.\footnote{We set $\alpha=4$ throughout to ensure a high level of sparsity.} By adjusting $\alpha$, the model effectively filters out noise, allowing the target token to be represented as a linear combination of only a few `core synonyms' with clear semantic correspondences. Accordingly, the initial embedding $\mathbf{e}^{t}_{x_t} \in \mathbb{R}^{d}$ is computed as follows:
\begin{equation}
\mathbf{e}^{t}_{x_t} = \sum_{i=1}^{N} p_{x_{ti}} \, \mathbf{e}^{s}_{x_{si}}.
\end{equation}
This approach preserves the embedding dimension $d$ and ensures immediate compatibility with the source model without requiring any architectural modifications. Through this process, all tokens in the set $\mathcal{R}$ are precisely initialized by being mapped to their respective optimal positions within the semantic space learned by the source model. Consequently, the expansion to the target language is achieved in a manner that inherits the source model's sparse encoding capability without loss.

\begin{table*}[t]
\centering

\renewcommand{\arraystretch}{1.0}

\setlength{\tabcolsep}{7pt} 

\resizebox{0.9\textwidth}{!}{%
\begin{tabular}{@{}ll *{8}{c} | *{8}{c}@{}} 
\toprule
\multirow{2}{*}{\textbf{Model}} & \multirow{2}{*}{\textbf{Lang.}} & \multicolumn{8}{c|}{\textbf{WebFAQ}} & \multicolumn{8}{c}{\textbf{MIRACL}} \\
\cmidrule(lr){3-10} \cmidrule(l){11-18}
& & Base & Rand. & Mean & Univ. & Multiv. & Focus & Ofa & SemB. & Base & Rand. & Mean & Univ. & Multiv. & Focus & Ofa & SemB. \\
\midrule

\multirow{6}{*}{Splade-v3} 
  & Ar  & 0.001 & 0.000 & 0.074 & 0.000 & 0.313 & 0.216 & 0.312 & \textbf{0.365} & 0.001 & 0.000 & 0.000 & 0.000 & 0.146 & 0.046 & 0.171 & \textbf{0.230} \\
  & Zh  & 0.006 & 0.001 & 0.134 & 0.001 & 0.438 & 0.200 & 0.386 & \textbf{0.545} & 0.006 & 0.000 & 0.002 & 0.000 & 0.035 & 0.005 & 0.036 & \textbf{0.101} \\
  & Hi  & 0.008 & 0.068 & 0.001 & 0.312 & 0.291 & 0.278 & 0.377 & \textbf{0.377} & 0.008 & 0.000 & 0.000 & 0.000 & 0.074 & 0.098 & 0.062 & \textbf{0.165} \\
  & Ko  & 0.028 & 0.000 & 0.089 & 0.001 & 0.421 & 0.115 & 0.482 & \textbf{0.520} & 0.028 & 0.000 & 0.005 & 0.000 & 0.172 & 0.013 & 0.196 & \textbf{0.236} \\
  & Ru  & 0.001 & 0.000 & 0.049 & 0.000 & 0.206 & 0.073 & 0.197 & \textbf{0.301} & 0.001 & 0.000 & 0.005 & 0.000 & 0.093 & 0.013 & 0.087 & \textbf{0.162} \\
  \cmidrule{2-18}
  & \textit{Avg.} & 0.009 & 0.014 & 0.069 & 0.063 & 0.334 & 0.176 & 0.351 & \textbf{0.422} & 0.009 & 0.000 & 0.002 & 0.000 & 0.104 & 0.035 & 0.110 & \textbf{0.179} \\
\midrule

\multirow{6}{*}{Splade-PP} 
  & Ar  & 0.001 & 0.000 & 0.090 & 0.002 & 0.302 & 0.099 & 0.293 & \textbf{0.328} & 0.001 & 0.000 & 0.000 & 0.000 & \textbf{0.162} & 0.001 & 0.160 & 0.161 \\
  & Zh  & 0.002 & 0.000 & 0.162 & 0.043 & 0.343 & 0.206 & 0.361 & \textbf{0.549} & 0.002 & 0.000 & 0.000 & 0.007 & 0.036 & 0.004 & 0.031 & \textbf{0.076} \\
  & Hi  & 0.004 & 0.079 & 0.010 & 0.366 & 0.275 & 0.352 & 0.377 & \textbf{0.377} & 0.004 & 0.000 & 0.000 & 0.000 & 0.075 & 0.081 & 0.054 & \textbf{0.125} \\
  & Ko  & 0.022 & 0.000 & 0.098 & 0.009 & 0.468 & 0.095 & 0.436 & \textbf{0.499} & 0.022 & 0.000 & 0.006 & 0.000 & \textbf{0.213} & 0.016 & 0.167 & 0.189 \\
  & Ru  & 0.004 & 0.000 & 0.058 & 0.001 & 0.195 & 0.073 & 0.212 & \textbf{0.290} & 0.004 & 0.000 & 0.005 & 0.000 & 0.107 & 0.010 & 0.094 & \textbf{0.141} \\
  \cmidrule{2-18}
  & \textit{Avg.} & 0.007 & 0.016 & 0.084 & 0.084 & 0.317 & 0.165 & 0.336 & \textbf{0.409} & 0.007 & 0.000 & 0.002 & 0.001 & 0.119 & 0.022 & 0.101 & \textbf{0.138} \\
\midrule

\multirow{6}{*}{Opensearch-Sparse-v1} 
  & Ar  & 0.000 & 0.000 & 0.078 & 0.000 & 0.348 & 0.092 & 0.301 & \textbf{0.317} & 0.000 & 0.000 & 0.000 & 0.000 & \textbf{0.221} & 0.001 & 0.182 & 0.151 \\
  & Zh  & 0.001 & 0.000 & 0.135 & 0.001 & 0.445 & 0.123 & 0.423 & \textbf{0.541} & 0.001 & 0.000 & 0.002 & 0.000 & 0.031 & 0.000 & 0.030 & \textbf{0.076} \\
  & Hi  & 0.004 & 0.075 & 0.000 & 0.398 & 0.298 & 0.352 & 0.359 & \textbf{0.359} & 0.004 & 0.000 & 0.000 & 0.000 & 0.103 & 0.082 & 0.078 & \textbf{0.134} \\
  & Ko  & 0.011 & 0.000 & 0.075 & 0.000 & 0.465 & 0.085 & \textbf{0.481} & 0.444 & 0.011 & 0.000 & 0.002 & 0.000 & \textbf{0.237} & 0.012 & 0.203 & 0.155 \\
  & Ru  & 0.003 & 0.000 & 0.042 & 0.000 & 0.185 & 0.059 & 0.215 & \textbf{0.271} & 0.003 & 0.000 & 0.004 & 0.000 & 0.112 & 0.013 & 0.108 & \textbf{0.120} \\
  \cmidrule{2-18}
  & \textit{Avg.} & 0.004 & 0.015 & 0.066 & 0.080 & 0.348 & 0.142 & 0.356 & \textbf{0.386} & 0.004 & 0.000 & 0.002 & 0.000 & \textbf{0.141} & 0.022 & 0.120 & 0.127 \\
\midrule

\multirow{6}{*}{Granite-30M-Sparse} 
  & Ar  & 0.002 & 0.000 & 0.038 & 0.000 & 0.000 & 0.044 & 0.000 & \textbf{0.158} & 0.002 & 0.000 & 0.000 & 0.000 & 0.000 & 0.001 & 0.000 & \textbf{0.061} \\
  & Zh  & 0.000 & 0.000 & 0.140 & 0.000 & 0.001 & 0.004 & 0.000 & \textbf{0.534} & 0.000 & 0.000 & 0.000 & 0.000 & 0.000 & 0.000 & 0.000 & \textbf{0.108} \\
  & Hi  & 0.002 & 0.000 & 0.070 & 0.000 & 0.000 & 0.063 & 0.000 & \textbf{0.148} & 0.002 & 0.000 & 0.000 & 0.000 & 0.000 & 0.015 & 0.001 & \textbf{0.042} \\
  & Ko  & 0.037 & 0.000 & 0.096 & 0.000 & 0.002 & 0.079 & 0.001 & \textbf{0.010} & 0.037 & 0.000 & 0.002 & 0.000 & 0.000 & 0.005 & 0.000 & 0.000 \\
  & Ru  & 0.008 & 0.000 & 0.026 & 0.000 & 0.000 & 0.003 & 0.000 & \textbf{0.060} & 0.008 & 0.000 & 0.002 & 0.000 & 0.000 & 0.004 & 0.000 & \textbf{0.028} \\
  \cmidrule{2-18}
  & \textit{Avg.} & 0.010 & 0.000 & 0.074 & 0.000 & 0.001 & 0.039 & 0.000 & \textbf{0.182} & 0.010 & 0.000 & 0.001 & 0.000 & 0.000 & 0.005 & 0.000 & \textbf{0.048} \\
\bottomrule
\end{tabular}%
}
\caption{Overall zero-shot retrieval performance comparison (nDCG@10) of various initialization methodologies. Comparison across diverse sparse encoders and language pairs on WebFAQ and MIRACL.}
\label{tab:zeroshot}

\end{table*}

\section{Experimental Setup}
\subsection{Training}

We use four sparse encoders: splade-v3~\cite{spladev3}, Splade\_PP\_en\_v1~\cite{spladepp}, opensearch-neural-sparse-encoding-v1\footnote{\url{https://huggingface.co/opensearch-project/opensearch-neural-sparse-encoding-v1}}, and granite-embedding-30m-sparse~\cite{granite}. For the target-language tokenizers, we use \texttt{ARBERT}~\cite{arabic} (Arabic), \texttt{bart-base-chinese}~\cite{chinese} (Chinese), \texttt{hindi-bert-v2}~\cite{hindi} (Hindi), \texttt{kobigbird-bert-base} (Korean), and \texttt{rubert-base-cased}~\cite{russian} (Russian). We fine-tune the transferred models independently for each target language using language-specific query-positive pairs from the multilingual WebFAQ~\cite{webfaq} dataset: Arabic (132k), Chinese (122k), Hindi (90k), Korean (92k), and Russian (377k). The training objective combines InfoNCE loss with a FLOPs regularization loss to enforce sparsity~\cite{spladev2}, using in-batch negatives for ranking. Detailed hyperparameters and hardware settings are provided in Appendix~\ref{training_details}.

\subsection{Evaluation} 
To quantitatively evaluate the retrieval performance of sparse retrieval models after transferring them to target languages, we utilize the evaluation sets of MIRACL~\cite{miracl} and WebFAQ~\cite{webfaq} across five languages: Arabic, Chinese, Hindi, Korean, and Russian. We adopt nDCG@10 as the primary retrieval performance metric and report FLOPS~\cite{splade} to assess the sparsity and efficiency.

\subsection{Baselines}
We compare SemBridge with several baseline methods for initializing the token embeddings of the target language tokenizer. For all approaches, embeddings of overlapping tokens are directly copied from the source embeddings without modification. For non-overlapping tokens, we compare our method against two categories of initialization strategies: (1) \textbf{Generic and Statistical Methods}: standard initialization techniques that do not explicitly model cross-lingual semantic correspondences, including Random, Mean, Univariate Gaussian, and Multivariate Gaussian. (2) \textbf{Language Transfer Methods}: methods for cross-lingual embedding initialization, specifically FOCUS~\cite{focus} and OFA~\cite{ofa}. Detailed formulations and descriptions of each baseline are provided in Appendix~\ref{appendix_baselines}.


\begin{table*}[t]
\centering

\renewcommand{\arraystretch}{1.0}

\setlength{\tabcolsep}{7pt} 

\resizebox{0.9\textwidth}{!}{%
\begin{tabular}{@{}ll *{8}{c} | *{8}{c}@{}} 
\toprule
\multirow{2}{*}{\textbf{Model}} & \multirow{2}{*}{\textbf{Lang.}} & \multicolumn{8}{c|}{\textbf{WebFAQ}} & \multicolumn{8}{c}{\textbf{MIRACL}} \\
\cmidrule(lr){3-10} \cmidrule(l){11-18}
& & Base & Rand. & Mean & Univ. & Multiv. & Focus & Ofa & SemB. & Base & Rand. & Mean & Univ. & Multiv. & Focus & Ofa & SemB. \\
\midrule

\multirow{6}{*}{Splade-v3} 
  & Ar  & 0.630 & 0.434 & 0.136 & 0.600 & 0.593 & 0.595 & 0.596 & \textbf{0.655} & 0.224 & 0.116 & 0.001 & 0.393 & 0.405 & 0.246 & 0.408 & \textbf{0.423} \\
  & Zh  & 0.442 & 0.644 & 0.420 & 0.759 & 0.765 & 0.704 & 0.746 & \textbf{0.774} & 0.019 & 0.027 & 0.017 & 0.129 & 0.185 & 0.100 & 0.164 & \textbf{0.203} \\
  & Hi  & 0.625 & 0.002 & 0.151 & 0.693 & 0.691 & 0.699 & 0.690 & \textbf{0.728} & 0.161 & 0.000 & 0.000 & 0.294 & 0.293 & 0.268 & 0.304 & \textbf{0.322} \\
  & Ko  & 0.606 & 0.563 & 0.178 & 0.750 & 0.755 & 0.671 & 0.742 & \textbf{0.762} & 0.237 & 0.180 & 0.023 & 0.416 & 0.428 & 0.261 & 0.362 & \textbf{0.382} \\
  & Ru  & 0.516 & 0.333 & 0.141 & 0.419 & 0.418 & 0.452 & 0.424 & \textbf{0.567} & 0.249 & 0.079 & 0.012 & 0.236 & 0.226 & 0.174 & 0.220 & \textbf{0.267} \\
  \cmidrule{2-18}
  & \textit{Avg.} & 0.564 & 0.395 & 0.205 & 0.644 & 0.644 & 0.624 & 0.640 & \textbf{0.697} & 0.178 & 0.080 & 0.011 & 0.294 & 0.307 & 0.210 & 0.292 & \textbf{0.319} \\
\midrule

\multirow{6}{*}{Splade-PP} 
  & Ar  & 0.630 & 0.277 & 0.137 & 0.602 & 0.602 & 0.564 & 0.597 & \textbf{0.640} & 0.224 & 0.043 & 0.001 & 0.391 & 0.373 & 0.270 & 0.364 & \textbf{0.386} \\
  & Zh  & 0.431 & 0.596 & 0.419 & 0.759 & 0.752 & 0.697 & 0.742 & \textbf{0.763} & 0.009 & 0.029 & 0.011 & 0.126 & 0.115 & 0.095 & 0.093 & \textbf{0.158} \\
  & Hi  & 0.609 & 0.161 & 0.163 & 0.692 & 0.687 & 0.690 & 0.678 & \textbf{0.712} & 0.152 & 0.000 & 0.000 & 0.303 & 0.281 & 0.265 & 0.275 & \textbf{0.322} \\
  & Ko  & 0.646 & 0.543 & 0.184 & 0.751 & 0.749 & 0.678 & 0.740 & \textbf{0.756} & 0.231 & 0.134 & 0.024 & 0.402 & \textbf{0.401} & 0.284 & 0.354 & 0.343 \\
  & Ru  & 0.521 & 0.340 & 0.161 & 0.418 & 0.416 & 0.441 & 0.419 & \textbf{0.549} & 0.229 & 0.077 & 0.008 & 0.228 & 0.223 & 0.180 & 0.213 & \textbf{0.281} \\
  \cmidrule{2-18}
  & \textit{Avg.} & 0.567 & 0.383 & 0.213 & 0.644 & 0.641 & 0.614 & 0.635 & \textbf{0.684} & 0.169 & 0.057 & 0.009 & 0.290 & 0.279 & 0.219 & 0.260 & \textbf{0.298} \\
\midrule

\multirow{6}{*}{Opensearch-Sparse-v1} 
  & Ar  & 0.620 & 0.064 & 0.135 & 0.597 & 0.588 & 0.576 & 0.603 & \textbf{0.648} & 0.209 & 0.001 & 0.001 & 0.416 & 0.348 & 0.247 & \textbf{0.387} & 0.382 \\
  & Zh  & 0.445 & 0.083 & 0.433 & 0.754 & 0.761 & 0.703 & 0.746 & \textbf{0.767} & 0.018 & 0.000 & 0.018 & 0.116 & 0.128 & 0.093 & 0.115 & \textbf{0.169} \\
  & Hi  & 0.631 & 0.109 & 0.160 & 0.686 & 0.688 & 0.699 & 0.683 & \textbf{0.707} & 0.169 & 0.000 & 0.001 & 0.298 & 0.295 & 0.253 & 0.287 & \textbf{0.311} \\
  & Ko  & 0.638 & 0.456 & 0.176 & 0.744 & 0.746 & 0.669 & 0.735 & \textbf{0.760} & 0.248 & 0.046 & 0.031 & 0.398 & \textbf{0.422} & 0.239 & 0.361 & 0.383 \\
  & Ru  & 0.477 & 0.004 & 0.163 & 0.412 & 0.413 & 0.449 & 0.419 & \textbf{0.532} & 0.177 & 0.000 & 0.010 & 0.228 & 0.229 & 0.166 & 0.221 & \textbf{0.234} \\
  \cmidrule{2-18}
  & \textit{Avg.} & 0.562 & 0.143 & 0.213 & 0.639 & 0.639 & 0.619 & 0.637 & \textbf{0.683} & 0.164 & 0.009 & 0.012 & 0.291 & 0.284 & 0.200 & 0.274 & \textbf{0.296} \\
\midrule

\multirow{6}{*}{Granite-30M-Sparse} 
  & Ar  & 0.562 & 0.219 & 0.119 & 0.000 & 0.420 & 0.512 & 0.435 & \textbf{0.623} & 0.172 & 0.016 & 0.000 & 0.000 & 0.209 & 0.203 & 0.226 & \textbf{0.301} \\
  & Zh  & 0.685 & 0.465 & 0.296 & 0.000 & 0.691 & 0.676 & 0.697 & \textbf{0.760} & 0.055 & 0.008 & 0.001 & 0.000 & 0.058 & 0.062 & 0.055 & \textbf{0.136} \\
  & Hi  & 0.574 & 0.156 & 0.131 & 0.000 & 0.504 & 0.671 & 0.535 & \textbf{0.706} & 0.106 & 0.001 & 0.001 & 0.000 & 0.049 & 0.240 & 0.068 & \textbf{0.249} \\
  & Ko  & 0.671 & 0.394 & 0.179 & 0.000 & 0.662 & 0.629 & 0.671 & \textbf{0.717} & 0.253 & 0.066 & 0.032 & 0.000 & 0.267 & 0.243 & 0.266 & \textbf{0.289} \\
  & Ru  & 0.469 & 0.251 & 0.112 & 0.000 & 0.330 & 0.398 & 0.331 & \textbf{0.526} & 0.211 & 0.050 & 0.007 & 0.000 & 0.105 & 0.148 & 0.096 & \textbf{0.250} \\
  \cmidrule{2-18}
  & \textit{Avg.} & 0.592 & 0.297 & 0.167 & 0.000 & 0.521 & 0.577 & 0.534 & \textbf{0.666} & 0.160 & 0.028 & 0.008 & 0.000 & 0.138 & 0.179 & 0.142 & \textbf{0.245} \\
\bottomrule
\end{tabular}%
}
\caption{Overall fine-tuned retrieval performance comparison (nDCG@10) of various initialization methodologies. Comparison across diverse sparse encoders and language pairs on WebFAQ and MIRACL.}
\label{tab:fine-tuning}

\end{table*}

\begin{figure*}[!t]
    \centering   
    \includegraphics[width=0.90\textwidth]{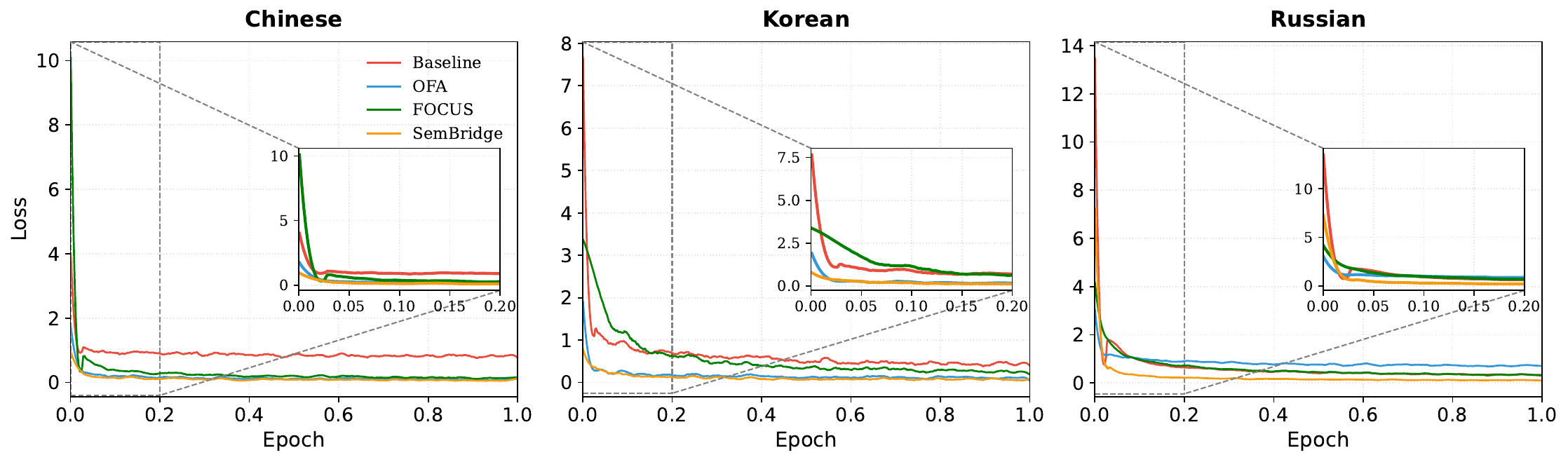}
    \caption{Training loss trajectories for Splade-v3 on Chinese, Korean, and Russian using Baseline, OFA, FOCUS, and SemBridge. The $y$-axis is Loss and the $x$-axis is Epoch. Insets zoom in on the initial training stage (0-0.2 epoch)}
\label{fig:loss}
\end{figure*}

\section{Experimental Results}
\subsection{Zero-shot Language Transfer}
Table~\ref{tab:zeroshot} presents the retrieval performance immediately following various embedding initialization strategies, illustrating how effectively each method transfers the source model's semantic knowledge inherent in the embedding layer to the target language. Here, Base denotes the original sparse encoder without tokenizer replacement or alignment. The results show that the Base model, which lacks an alignment process, along with simple statistical approaches such as Random and Mean, yields near-zero or marginal performance across most language pairs. While univariate (Univar.) and multivariate (Multivar.) lead to limited improvements in certain settings, they exhibit high variance across languages and suffer from sharp performance degradation depending on the model architecture.

In contrast, SemBridge consistently demonstrates superior initialization performance across all four models. Notably, it records average zero-shot scores of 0.422 and 0.522 for Splade-v3 and Splade-PP, respectively, on the WebFAQ dataset. This suggests that SemBridge effectively captures cross-lingual semantic correspondences within the representation space. These results substantiate the exceptional language transfer effectiveness of SemBridge, showing that it successfully transfers the source-language sparse encoder's capabilities to the target language and provides a strong starting point for fine-tuning while achieving high zero-shot retrieval performance.

\subsection{Impact of Initialization on Fine-tuning}
Table \ref{tab:fine-tuning} presents the results of subsequent fine-tuning using language-specific retrieval data after the initialization phase. In the majority of experimental settings, SemBridge consistently achieves superior performance, outperforming the baselines across five target languages, four models, and two datasets. For instance, with the Granite-30M-Sparse model, SemBridge was the sole method to surpass all baseline methods on both datasets. 

These results show that the effect of initialization is not limited to zero-shot transfer, but continues to influence the model after fine-tuning. Crucially, significant performance disparities persist even after applying an identical fine-tuning process, demonstrating that the quality of initialization fundamentally constrains the model's capabilities. This helps explain why existing methods can continue to fall behind after fine-tuning when their initial token correspondences are noisy or imprecise, as further supported by the qualitative analysis in Section~\ref{Qualititive}. In contrast, SemBridge provides a robust foundation for preserving and leveraging the source model's retrieval capabilities in the target language.

\begin{figure*}[t!]
    \centering
    \resizebox{0.9\textwidth}{!}{%
    \begin{minipage}{\textwidth}
        \centering
        \begin{minipage}[c]{0.03\textwidth}
            \centering
            \textbf{(a)}
        \end{minipage}%
        \begin{minipage}[c]{0.97\textwidth}
            \centering
            \includegraphics[width=\linewidth]{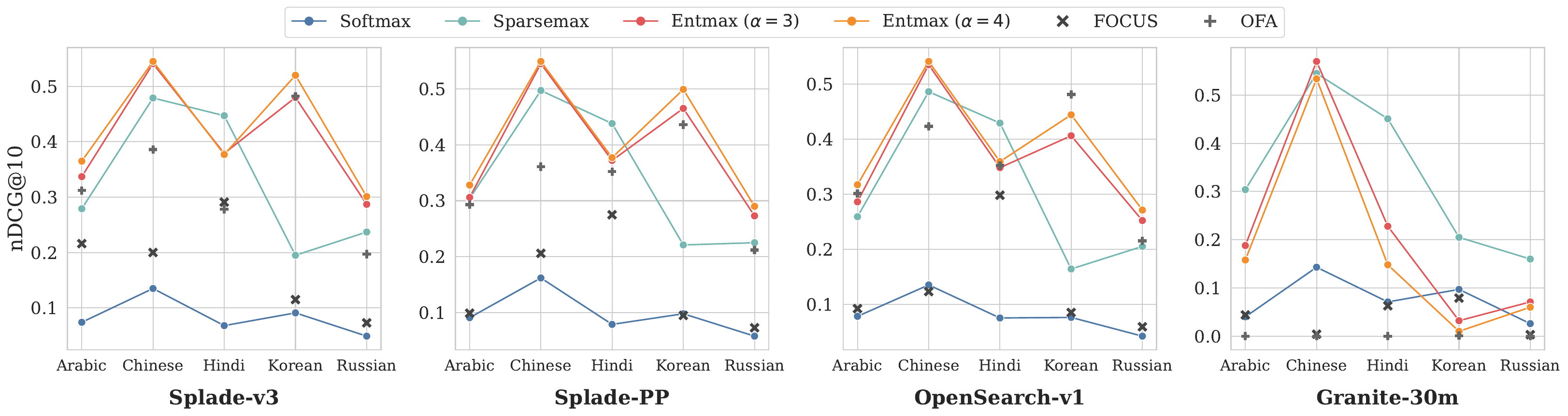}
        \end{minipage}

        \begin{minipage}[c]{0.03\textwidth}
            \centering
            \raisebox{1.5em}{\textbf{(b)}} 
        \end{minipage}%
        \begin{minipage}[c]{0.97\textwidth}
            \centering
            \includegraphics[width=\linewidth]{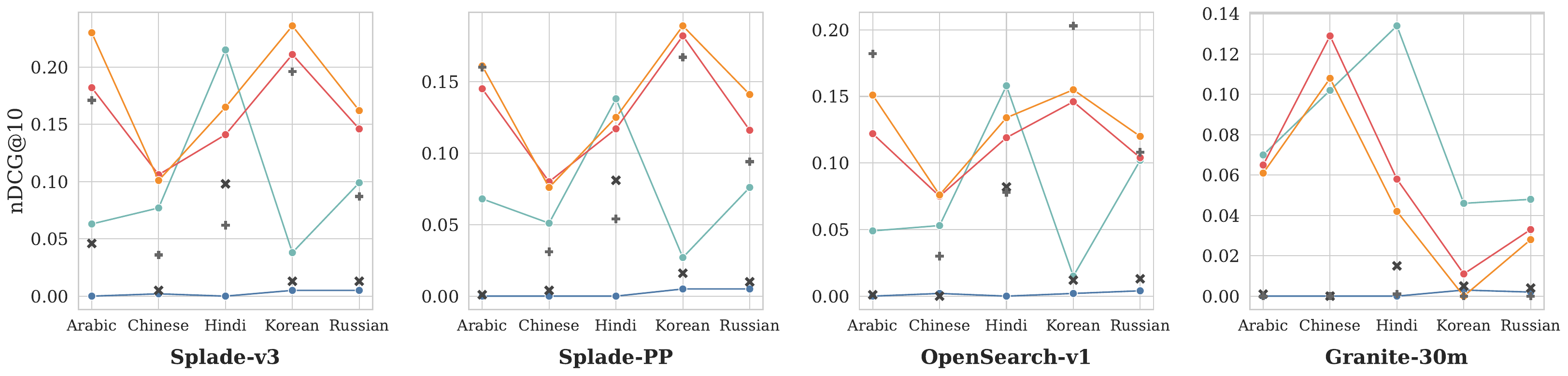}
        \end{minipage}
    \end{minipage}%
    }
    \caption{Zero-shot retrieval performance (nDCG@10) on (a) WebFAQ and (b) MIRACL. For SemBridge, the sparsity level $\alpha$ is varied from 1 to 4, with FOCUS and OFA included as baselines.}
    \label{fig:ablation}
\end{figure*}

\subsection{Loss Trajectory}
Figure~\ref{fig:loss} illustrates the training loss trajectories for the SPLADE-v3 model across various initialization methods. Each subplot displays the loss curves during training for the Baseline, OFA, FOCUS, and the proposed SemBridge in Chinese, Korean, and Russian. Experimental results reveal that SemBridge generally begins training with a significantly lower initial loss compared to other approaches.

It is worth noting that while the initial loss for Russian is slightly higher, it exhibits an immediate convergence pattern. This suggests that our initialization strategy provides an optimal initial embedding state for the model, enabling it to swiftly converge to a stable position within the loss landscape. Furthermore, SemBridge demonstrates exceptional efficiency with a steep decline in loss during the early stages of training. This rapid adaptability is a crucial factor that allows the model to quickly learn target-language characteristics even under constrained resources. Consequently, SemBridge maintains the lowest loss throughout the training process, achieving a superior representation quality upon final convergence compared to both the Baseline and other competitive methods. While all methods exhibit stable convergence curves, SemBridge stands out across all metrics, including initialization, training efficiency, and final performance, thereby empirically validating its effectiveness in preserving the source model’s capabilities for the target language.

\section{Ablation and Analysis}

\subsection{Analysis of Sparse Weighting}
To validate the effectiveness of similarity-based sparse weighting (Eq.~\eqref{eq:sparse_weighting}), Figure~\ref{fig:ablation} analyzes performance trends across varying Entmax hyperparameters ($\alpha \in \{1, 2, 3, 4\}$). Entmax generalizes softmax ($\alpha=1$) and sparsemax ($\alpha=2$), allowing us to examine how different sparsity levels affect transfer performance. The softmax case, which weights all source tokens, yields the lowest scores across most settings. This suggests that semantically irrelevant source tokens introduce noise during target embedding initialization, disrupting precise semantic alignment. In contrast, sparse weighting methods, such as sparsemax ($\alpha=2$) and Entmax ($\alpha \ge 3$), substantially improve transfer performance by reducing interference from irrelevant semantic information. Specifically, configurations with $\alpha \ge 2$ demonstrate consistently high performance. Entmax ($\alpha=3, 4$) further outperforms the existing baselines, FOCUS and OFA, indicating that focusing on a few core tokens is effective for target embedding initialization. However, excessive sparsity may exclude meaningful semantic clues, implying that an appropriate $\alpha$ should be selected based on the target language. Overall, these results confirm that SemBridge maintains robustness across a range of sparsity settings and achieves superior performance under sparse configurations.

\begin{figure}[t!]
  \center
  \includegraphics[width=0.9\columnwidth]{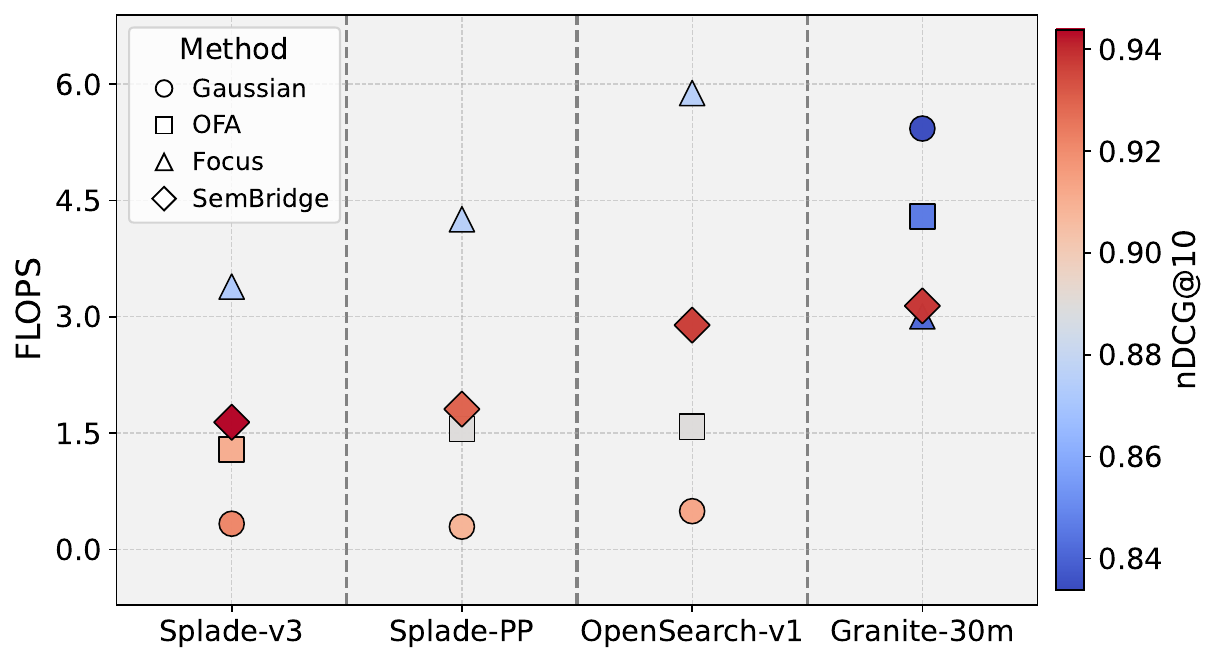}
  \caption{Efficiency vs. performance trade-off on the MIRACL dev set (Chinese). The y-axis shows FLOPS, and marker color indicates nDCG@10.}
  \label{fig:flops}
\end{figure}

\subsection{Efficiency Analysis}

\begin{table*}[t]
\centering
\includegraphics[width=1.0\textwidth]{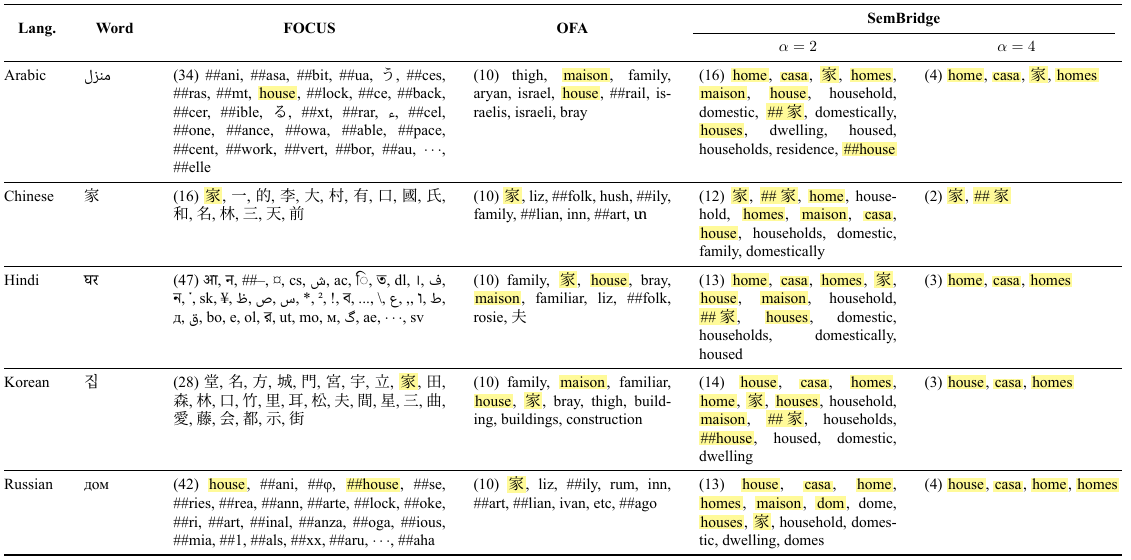}
\caption{Qualitative comparison of top-weighted source tokens for \texttt{splade-v3} initialization of target-language words corresponding to ``home'' across five languages. Tokens are listed in descending order of similarity for each method, and highlighted tokens indicate direct semantic equivalents and core matches for the concept of ``home''.}
\label{tab:qualititive}
\end{table*}

Figure~\ref{fig:flops} compares the efficiency and retrieval performance of different methods after fine-tuning. Experimental results reveal that FOCUS yields excessively high FLOPS. Conversely, Gaussian achieves very low FLOPS, but its performance is highly unstable, exhibiting severe degradation on Granite-30m. While OFA maintains moderate FLOPS, its performance improvement remains limited. SemBridge effectively balances computational efficiency and retrieval performance, yielding the highest nDCG@10 scores across all models while maintaining significantly lower FLOPS compared to FOCUS. This indicates that SemBridge successfully transfers semantic knowledge while maintaining a highly sparse representation. These results confirm that SemBridge provides a robust initialization strategy, ensuring high retrieval performance while maintaining computational efficiency.

\subsection{Qualitative Analysis}
\label{Qualititive}
Table~\ref{tab:qualititive} compares the top source tokens assigned to target language terms by each initialization strategy during target-language token initialization. For tokens representing `home' across five different languages, FOCUS mapped meaningless subwords such as `ani' and `\#\#asa' or contextually irrelevant tokens in Arabic, highlight the limitations of simple similarity-based approaches. OFA showed some improvement by capturing terms like `family' and `maison'; however, it still exhibited unstable semantic alignment, assigning unrelated words such as `israel' for Arabic inputs or `hush' for Chinese. These observations suggest that existing methodologies struggle to achieve precise cross-lingual semantic alignment and fail to fully transfer the source model's capability to the target language.

In contrast, SemBridge exhibits superior semantic consistency by accurately linking `home'-related tokens across all five languages to English terms like `home' and `house', as well as multilingual synonyms such as `casa' and `maison'. Specifically, the sparsity of these mappings can be controlled by adjusting the Entmax $\alpha$ value. Setting $\alpha$ = 2 captures broad contextual information such as `dwelling' and `households', whereas setting $\alpha$ = 4 effectively suppresses noise by selecting only core synonyms. These observations demonstrate that SemBridge effectively transfers language-agnostic semantic information during initialization.

\section{Conclusion}
In this paper, we proposed SemBridge, an embedding initialization method designed to transfer English-centric sparse encoders to target languages. By leveraging multilingual dense embeddings as a semantic bridge, SemBridge aligns source and target vocabularies. It applies a sparse weighting mechanism to initialize target embeddings from semantically relevant source tokens, effectively filtering out noise. Extensive experiments across five languages (Arabic, Chinese, Hindi, Korean, and Russian) and four sparse architectures show that SemBridge outperforms existing initialization methods in both zero-shot and fine-tuned settings. Furthermore, it significantly accelerates convergence during fine-tuning, demonstrating that accurate token-level semantic alignment is crucial for preserving and transferring sparse retrieval capabilities. 
Ultimately, our results establish SemBridge as a robust and practical solution for building high-performance sparse encoders in target languages.

\section*{Limitations}
Our study has the following limitations. First, although SemBridge demonstrates consistent effectiveness across the five target languages evaluated in this work (Arabic, Chinese, Hindi, Korean, and Russian), our evaluation remains limited to this language set. As a result, its behavior in languages with different resource levels, scripts, or morphological structures remains underexplored. Second, target embedding initialization depends on the sparsity level controlled by the Entmax hyperparameter $\alpha$. While SemBridge remains robust across several sparsity settings, the optimal value may vary by target language and tokenizer. Excessive sparsity may exclude meaningful semantic clues, whereas insufficient sparsity may introduce irrelevant source-token noise. Investigating adaptive or language-aware sparsity selection remains an important direction for future work.

\section*{Ethical Considerations}
This work focuses on adapting sparse retrieval models to target-language environments. Our experiments use publicly available datasets and do not involve collecting personal information or interacting with human subjects. Since SemBridge relies on multilingual dense embedding models as semantic bridges, biases or uneven language coverage in these models may affect the initialized sparse encoder. Therefore, real-world use in new languages or domains should be accompanied by careful validation.

\bibliography{custom}

\clearpage
\appendix

\section{Token Distribution Analysis}
\label{app:token_distribution}
To examine the linguistic composition of the vocabulary in sparse encoders, we conduct the token distribution analysis presented in Figure~\ref{fig:teaser}. We analyze the four source sparse encoders used in our experiments: \texttt{splade-v3}, \texttt{Splade\_PP\_en\_v1}, \texttt{opensearch-neural-sparse-encoding-v1}, and \texttt{granite-embedding-30m-sparse}. The tokens in each model's vocabulary are categorized by language using a language detection library\footnote{\url{https://github.com/pemistahl/lingua-rs}}. We consider English and the five target languages used in our experiments: Arabic, Chinese, Hindi, Korean, and Russian. Tokens corresponding to other languages are grouped into \textit{Other}, while tokens without clear linguistic characters, such as special tokens, numbers, punctuation marks, and symbols, are grouped into \textit{ETC}.

Specifically, we do not directly use the raw token strings in the vocabulary. Instead, we decode each token ID and determine the language of the decoded token. This is because raw token strings may include tokenizer-specific subword prefixes or whitespace markers, which can interfere with identifying the language of the actual token. Therefore, we compute the token distribution based on decoded tokens. The analysis shows that the vocabularies of existing sparse encoders are generally concentrated on English tokens, while target-language tokens account for only a limited portion. These distributional differences arise because the vocabulary of each sparse encoder is determined by the tokenizer of the backbone model used for training. Specifically, \texttt{granite-embedding-30m-sparse} uses RoBERTa~\cite{roberta} as its backbone, whereas the other three models use BERT~\cite{bert} as their backbone. Thus, the observed token distribution reflects the vocabulary composition of each backbone tokenizer.

\section{Baselines}
\label{appendix_baselines}
We compare SemBridge with several baseline methods for initializing the token embeddings of the target language tokenizer. For all approaches, embeddings of overlapping tokens are directly copied from the source embeddings without modification. For non-overlapping tokens, different initialization strategies are applied as follows:

\header{Random.}
Each new token embedding $\mathbf{e}^{t}_{x_t}$ ($x_t \in \mathcal{R}$) is independently sampled from a normal distribution with a zero mean and small variance: $\mathbf{e}^{t}_{x_t} \sim \mathcal{N}(\mathbf{0}, \sigma^2 \mathbf{I})$, where $\sigma = 0.02$. While this is a standard initialization strategy, it does not leverage any semantic information from the source embeddings.

\header{Mean.}
All remaining token embeddings are initialized with the global mean of all source embeddings: $\mathbf{e}^{t}_{x_t} = \frac{1}{N} \sum_{i=1}^{N} \mathbf{e}^{s}_{x_{s_i}}$ ($\forall x_t \in \mathcal{R}$), where $N = |\mathcal{V}_{s}|$ denotes the size of the source vocabulary. In this approach, every new token shares the same initial value.

\header{Univariate.}
The remaining tokens $x_t$ are sampled from a univariate Gaussian distribution $\mathcal{N}(\mu_{g}, \sigma_{g}^{2})$ sharing global statistics of the source embedding matrix $\mathbf{E}^{s}$. Here, $\mu_{g}$ and $\sigma_{g}^{2}$ represent the mean and variance calculated across all elements in the source embeddings. While this strategy captures the overall scale of the source embedding space, it ignores dimension-specific characteristics.

\header{Multivariate.}
Each dimension $j$ of the token embedding is sampled from a diagonal multivariate Gaussian distribution: $\mathbf{e}^{t}_{x_t}[j] \sim \mathcal{N}(\mu_{j}, \sigma_{j}^{2})$, where $\mu_{j}$ and $\sigma_{j}^{2}$ denote the mean and variance of the $j$-th dimension across the source embeddings. This method reflects the statistical properties of the source embedding space more precisely by accounting for the varying scales and distributions across different dimensions.

\header{FOCUS~\cite{focus}.}
This approach computes the similarity between remaining tokens $x_t \in \mathcal{R}$ and overlapping tokens $x \in \mathcal{V}_o$ within an auxiliary static embedding space trained on a target corpus. Based on these similarities, the new token embeddings are initialized as a weighted average of the source embeddings of the overlapping tokens. While this method allows the transfer of the source model's semantic space without explicit cross-lingual alignment, it is strictly limited by the fact that the reference set for synthesis is restricted only to overlapping tokens $\mathcal{V}_o$. Consequently, it fails to exploit other semantically relevant source tokens that do not belong to the overlapping set.

\begin{table*}[t!]
\centering
\small
\setlength{\tabcolsep}{6pt}
\begin{tabular}{llcl}
\toprule
\textbf{Language} & \textbf{Model} & \textbf{Vocab Size} & \textbf{HF Repo.} \\
\midrule
Arabic  & \texttt{ARBERT} 
& 100,000 
& \url{https://huggingface.co/UBC-NLP/ARBERT} \\
Chinese & \texttt{bart-base-chinese} 
& 51,271 
& \url{https://huggingface.co/OpenMOSS-Team/bart-base-chinese} \\
Hindi   & \texttt{hindi-bert-v2} 
& 197,285 
& \url{https://huggingface.co/l3cube-pune/hindi-bert-v2} \\
Korean  & \texttt{kobigbird-bert-base} 
& 32,500 
& \url{https://huggingface.co/monologg/kobigbird-bert-base} \\
Russian & \texttt{rubert-base-cased} 
& 119,547 
& \url{https://huggingface.co/DeepPavlov/rubert-base-cased} \\
\bottomrule
\end{tabular}
\caption{Summary of the target-language tokenizers used in our experiments. The table reports the model, vocabulary size, and Hugging Face repository for each target language.}
\label{tab:target_tokenizers}
\end{table*}

\header{OFA~\cite{ofa}.}
This framework utilizes matrix factorization to initialize token embeddings by decomposing the source embedding matrix $\mathbf{E}^{s}$ into language-independent primitive embeddings $\mathbf{P}$ and token-specific coordinates $\mathbf{F}^{s}$. For the remaining tokens $x_t \in \mathcal{R}$, the initialization is performed by generating target coordinates $\mathbf{F}^{t}$ through a convex combination of source coordinates, using similarity weights derived from an external multilingual word vector space such as ColexNet+. The final synthesized coordinates are then projected onto the original dimensions via the primitive embeddings $\mathbf{P}$ to serve as the target embeddings.

\section{Experiments Details}
\subsection{Target Tokenizers}
\label{app:target_tokenizers}
Table~\ref{tab:target_tokenizers} presents the target-language tokenizers used in our experiments. To construct the target-language vocabulary, we use the vocabulary of a pretrained tokenizer for each language. Specifically, we replace the tokenizer of the source sparse encoder with the target tokenizer for each language, and define the corresponding tokenizer vocabulary as the target vocabulary $V_t$ for target token embedding initialization. Based on the target vocabulary $V_t$, SemBridge initializes the embedding layer of the source sparse encoder after tokenizer replacement, enabling the model to be transferred to the target language.

\subsection{Hyperparams \& Hardware}
\label{training_details}
All fine-tuning is conducted on four NVIDIA A100 GPUs. The models are trained for 1 epoch with a total batch size of 64, a maximum sequence length of 512, and bf16 precision. We employ the AdamW optimizer with a learning rate of $2 \times 10^{-5}$ and a linear learning rate warm-up for 5\% of the total steps. For the training objective, the FLOPs regularization weights for documents and queries are set to $1 \times 10^{-4}$ and $3 \times 10^{-4}$, respectively. To ensure reproducibility, all random seeds, including the data shuffling seed, are fixed to 42.

\begin{table*}[t!]
\centering
\resizebox{0.85\textwidth}{!}{%
\begin{tabular}{@{} l l *{6}{c}|| *{6}{c} @{}} 
\toprule
\multirow{2}{*}{\textbf{Sparse Model}} & \multirow{2}{*}{\textbf{Method}} & \multicolumn{6}{c}{\textbf{WebFAQ}} & \multicolumn{6}{c}{\textbf{MIRACL}} \\
\cmidrule(lr){3-8} \cmidrule(lr){9-14}

 & & \textbf{Ar} & \textbf{Zh} & \textbf{Hi} & \textbf{Ko} & \textbf{Ru} & \textbf{Avg} & \textbf{Ar} & \textbf{Zh} & \textbf{Hi} & \textbf{Ko} & \textbf{Ru} & \textbf{Avg} \\
\midrule

\multirow{6}{*}{splade-v3} 
 & FOCUS 
 & 0.216 & 0.200 & 0.278 & 0.115 & 0.073 & 0.176 
 & 0.046 & 0.005 & 0.098 & 0.013 & 0.013 & 0.035 \\

 & OFA 
 & 0.312 & 0.386 & 0.377 & 0.482 & 0.197 & 0.351 
 & 0.171 & 0.036 & 0.062 & 0.196 & 0.087 & 0.110 \\
 \cmidrule(l){2-14}

 & \textbf{SemBridge}
 & 0.365 & 0.545 & 0.377 & 0.520 & 0.301 & \textbf{0.422} 
 & 0.230 & 0.101 & 0.165 & 0.236 & 0.162 & \textbf{0.179} \\
 \cdashline{2-14}[0.8pt/1.2pt]
 & \phantom{\textbf{@}} w/ MiniLM
 & 0.215 & 0.397 & 0.354 & 0.359 & 0.191 & 0.303 
 & 0.089 & 0.030 & 0.123 & 0.127 & 0.090 & 0.092 \\
 
 & \phantom{\textbf{@}} w/ mGTE
 & 0.348 & 0.471 & 0.393 & 0.384 & 0.293 & 0.378 
 & 0.206 & 0.061 & 0.167 & 0.157 & 0.173 & 0.148 \\
 
 & \phantom{\textbf{@}} w/ Qwen3
 & 0.288 & 0.313 & 0.324 & 0.389 & 0.215 & 0.306 
 & 0.145 & 0.026 & 0.105 & 0.183 & 0.101 & 0.114 \\
\midrule

\multirow{6}{*}{Splade\_PP\_en\_v1} 
 & FOCUS 
 & 0.099 & 0.206 & 0.352 & 0.095 & 0.073 & 0.165 
 & 0.001 & 0.004 & 0.081 & 0.016 & 0.010 & 0.022 \\

 & OFA 
 & 0.293 & 0.361 & 0.377 & 0.436 & 0.212 & 0.336 
 & 0.160 & 0.031 & 0.054 & 0.167 & 0.094 & 0.101 \\
 \cmidrule(l){2-14}

 & \textbf{SemBridge}
 & 0.328 & 0.549 & 0.377 & 0.499 & 0.290 & \textbf{0.409} 
 & 0.161 & 0.076 & 0.125 & 0.189 & 0.141 & \textbf{0.138} \\
 \cdashline{2-14}[0.8pt/1.2pt]
 & \phantom{\textbf{@}} w/ MiniLM
 & 0.166 & 0.355 & 0.317 & 0.311 & 0.176 & 0.265 
 & 0.055 & 0.016 & 0.089 & 0.078 & 0.071 & 0.062 \\

 & \phantom{\textbf{@}} w/ mGTE
 & 0.332 & 0.477 & 0.407 & 0.363 & 0.286 & 0.373 
 & 0.164 & 0.042 & 0.120 & 0.104 & 0.143 & 0.115 \\

 & \phantom{\textbf{@}} w/ Qwen3
 & 0.259 & 0.298 & 0.292 & 0.378 & 0.204 & 0.286 
 & 0.109 & 0.017 & 0.063 & 0.142 & 0.074 & 0.081 \\
\midrule

\multirow{6}{*}{\shortstack[l]{opensearch-neural\\-sparse-encoding-v1}} 
 & FOCUS 
 & 0.092 & 0.123 & 0.352 & 0.085 & 0.059 & 0.142 
 & 0.001 & 0.000 & 0.082 & 0.012 & 0.013 & 0.022 \\

 & OFA 
 & 0.301 & 0.423 & 0.359 & 0.481 & 0.215 & 0.356 
 & 0.182 & 0.030 & 0.078 & 0.203 & 0.108 & 0.120 \\
 \cmidrule(l){2-14}

 & \textbf{SemBridge}
 & 0.317 & 0.541 & 0.359 & 0.444 & 0.271 & \textbf{0.387} 
 & 0.151 & 0.076 & 0.134 & 0.155 & 0.120 & \textbf{0.127} \\
 \cdashline{2-14}[0.8pt/1.2pt]
 & \phantom{\textbf{@}} w/ MiniLM
 & 0.157 & 0.370 & 0.313 & 0.251 & 0.155 & 0.249 
 & 0.054 & 0.021 & 0.097 & 0.102 & 0.068 & 0.068 \\
 
 & \phantom{\textbf{@}} w/ mGTE
 & 0.313 & 0.475 & 0.389 & 0.244 & 0.263 & 0.337 
 & 0.155 & 0.035 & 0.134 & 0.110 & 0.130 & 0.113 \\
 
 & \phantom{\textbf{@}} w/ Qwen3
 & 0.238 & 0.229 & 0.275 & 0.328 & 0.167 & 0.247 
 & 0.094 & 0.014 & 0.072 & 0.156 & 0.067 & 0.081 \\
\midrule

\multirow{6}{*}{\shortstack[l]{granite-embedding\\-30m-sparse}} 
 & FOCUS 
 & 0.044 & 0.004 & 0.063 & 0.079 & 0.003 & 0.039 
 & 0.001 & 0.000 & 0.015 & 0.005 & 0.004 & 0.005 \\

 & OFA 
 & 0.000 & 0.000 & 0.000 & 0.001 & 0.000 & 0.000 
 & 0.000 & 0.000 & 0.001 & 0.000 & 0.000 & 0.000 \\
 \cmidrule(l){2-14}

 & \textbf{SemBridge}
 & 0.158 & 0.534 & 0.148 & 0.010 & 0.060 & 0.182 
 & 0.061 & 0.108 & 0.042 & 0.000 & 0.028 & 0.048 \\
 \cdashline{2-14}[0.8pt/1.2pt]
 & \phantom{\textbf{@}} w/ MiniLM
 & 0.121 & 0.404 & 0.199 & 0.044 & 0.060 & 0.166 
 & 0.041 & 0.023 & 0.047 & 0.007 & 0.023 & 0.028 \\
 
 & \phantom{\textbf{@}} w/ mGTE
 & 0.155 & 0.534 & 0.265 & 0.134 & 0.131 & \textbf{0.244} 
 & 0.072 & 0.121 & 0.067 & 0.039 & 0.070 & \textbf{0.074} \\
 
 & \phantom{\textbf{@}} w/ Qwen3
 & 0.169 & 0.056 & 0.268 & 0.153 & 0.121 & 0.153 
 & 0.065 & 0.000 & 0.059 & 0.019 & 0.046 & 0.038 \\
\bottomrule
\end{tabular}%
}
\caption{Impact of various dense bridge models on zero-shot retrieval performance (nDCG@10). We compare SemBridge utilizing different auxiliary models against existing initialization baselines (FOCUS, OFA) on WebFAQ and MIRACL datasets.}
\label{tab:bridge_change}

\end{table*}

\section{Additional Experiments}
\subsection{Robustness to Bridge Models}
To investigate the impact of the Cross-lingual Semantic Bridge ($\mathcal{B}$) on the transfer performance, we conduct comparative experiments using three additional multilingual embedding models with varying characteristics: paraphrase-multilingual-MiniLM-L12-v2~\cite{minilm}, gte-multilingual-base~\cite{gte}, and Qwen3-Embedding-0.6B~\cite{qwen3}. The resulting zero-shot retrieval performance is presented in Table~\ref{tab:bridge_change}.

The experimental results demonstrate that the choice of the bridge model influences the language transfer outcomes of sparse models. Specifically, we observe that knowledge transfer to the target language becomes more effective as more powerful dense models, such as mGTE or Qwen3, are utilized. For instance, in the Splade-v3 model on the WebFAQ dataset, the configuration using bge-m3 achieved the highest average performance of 0.422, followed by mGTE (0.378), Qwen3 (0.306), and MiniLM (0.303). These findings suggest that the refinement level of the cross-lingual alignment within the bridge model's semantic space is a key factor in determining the quality of initialization for the target language. This implies that the quality of the similarity matrix generated by the bridge model is a critical determinant of semantic alignment accuracy between source and target tokens. Crucially, however, SemBridge demonstrates stable and superior performance compared to existing methods, FOCUS and OFA, regardless of the specific bridge model employed. These results substantiate the robust generalizability of our proposed method. While performance scales with the bridge model's capacity, SemBridge maintains consistent robustness, effectively performing cross-lingual semantic mapping by leveraging the intrinsic structure of any given dense embedding space.

\end{document}